\documentclass[letterpaper]{article}

\usepackage{authblk}
\usepackage[style=ieee]{biblatex}
\usepackage[letterpaper, margin=1in]{geometry}
\usepackage{graphicx}
\usepackage{hyperref}
\usepackage{listings}
\usepackage{xcolor}

\hypersetup{
    colorlinks=True,
    urlcolor=blue,
    linkcolor=black,
    citecolor=black
}

\definecolor{codegreen}{rgb}{0,0.6,0}
\definecolor{codered}{rgb}{0.85,0,0.05}
\definecolor{backcolour}{rgb}{0.95,0.95,0.95}
\definecolor{codeblue}{rgb}{0.45,0.4,0.95}

\lstdefinestyle{mystyle}{
    backgroundcolor=\color{backcolour},
    commentstyle=\color{codegreen},
    keywordstyle=\color{codered},
    numberstyle=\tiny\color{gray},
    stringstyle=\color{codeblue},
    basicstyle=\sffamily\footnotesize,
    breakatwhitespace=true,
    breaklines=true,
    captionpos=b,
    keepspaces=true,
    numbersep=5pt,
    showspaces=false,
    showstringspaces=false,
    showtabs=false,
    tabsize=4
}
\title{\texttt{pyscreener}: A Python Wrapper for Computational Docking Software}

\author[1,2]{David E. Graff}
\author[2,3]{Connor W. Coley}
\affil[1]{Department of Chemistry and Chemical Biology, Harvard University, Cambridge, MA}
\affil[2]{Department of Chemical Engineering, Massachusetts Institute of Technology, Cambridge, MA}
\affil[3]{Department of Electrical Engineering and Computer Science, Massachusetts Institute of Technology, MIT, Cambridge, MA}
\affil[ ]{\textit{ccoley@mit.edu}}
\begin{document}

\maketitle

\section*{Summary}
\texttt{pyscreener} is a Python library that seeks to alleviate the challenges of large-scale structure-based design using computational docking. It provides a simple and uniform interface that is agnostic to the backend docking engine with which to calculate the docking score of a given molecule in a specified active site. Additionally, \texttt{pyscreener} features first-class support for task distribution, allowing users to seamlessly scale their code from a local, multi-core setup to a large, heterogeneous resource allocation.

\section*{Statement of Need}
Computational docking is an important technique in structure-based drug design that enables the rapid approximation of binding affinity for a candidate ligand in a matter of CPU seconds. With the growing popularity of ultra-large ligand libraries, docking is increasingly used to sift through hundreds of millions of compounds to try to identify novel and potent binders for a variety of protein targets \cite{gorgulla_open-source_2020,lyu_ultra-large_2019}.
There are many choices of docking software, and certain software are better suited towards specific protein-ligand contexts (e.g., flexible protein side chains or sugar-like ligand molecules). Switching between these software is often not trivial as the input preparation, simulation, and output parsing pipelines differ between each software.

In addition, many of these programs exist only as command-line applications and lack Python bindings.
This presents an additional challenge for their integration into molecular optimization workflows, such as reinforcement learning or genetic algorithms.
Molecular optimization objectives have largely been limited to benchmark tasks, such as penalized logP, QED, JNK3 or GSK3$\beta$ inhibitor classification \cite{li_multi-objective_2018}, and others contained in the GuacaMol library \cite{brown_guacamol_2019}. These benchmarks are useful for comparing molecular design techniques, but they are not representative of true drug discovery tasks in terms of complexity;
% While useful for comparing novel molecular design techniques, these benchmarks are not representative of true drug discovery tasks; 
computational docking is at least one step in the right direction.

% One challenge in this regard is that most docking programs accept input in the form of molecular supply files with predefined 3D geometry (e.g., Mol2 or PDBQT format), whereas many molecular optimization techniques propose new molecules in the form of SMILES strings
While many molecular optimization techniques propose new molecules in the form of SMILES strings \cite{elton_deep_2019}, most docking programs accept input in the form of
% However, these techniques generally propose new molecules in the form of SMILES strings \cite{elton_deep_2019}, but many docking programs accept input in the form of 
molecular supply files with predefined 3D geometry (e.g., Mol2 or PDBQT format).
Using the docking score of a molecule as a design objective thus requires an ad hoc implementation for which no standardized approach exists.
The \texttt{vina} library \cite{eberhardt_autodock_2021} is currently the only library capable of performing molecular docking within Python code, but it is limited to docking molecules using solely AutoDock Vina as the backend docking engine.
Moreover, the object model of the \texttt{vina} library accepts input ligands only as PDBQT files or strings and still does not address the need to quickly calculate the docking score of a molecule from its SMILES string.

In our work on the MolPAL software \cite{graff_accelerating_2021}, we required a library that is able to accept molecular inputs as SMILES strings and output their corresponding docking scores for a given receptor and docking box. Our use-case also called for docking large batches of molecules across large and distributed hardware setups. Lastly, we desired that our library be flexible with respect to the underlying docking engine, allowing us to use a variety of backend docking software (e.g., Vina \cite{trott_autodock_2010}, Smina \cite{koes_lessons_2013}, QVina \cite{alhossary_fast_2015}, or DOCK6 \cite{allen_dock_2015}) with minimal changes to client code.
To that end, we developed \texttt{pyscreener}, a Python library that is flexible with respect to both molecular input format and docking engine that transparently handles the distribution of docking simulations across large resource allocations.

\section*{Implementation and Performance}
The primary design goals with \texttt{pyscreener} were to (1) provide a simple interface with which to calculate the docking score of an input small molecule and (2) transparently distribute the corresponding docking simulations across a large resource allocation. The object model of \texttt{pyscreener} relies on four classes: \texttt{CalculationData}, \texttt{CalculationMetadata}, \texttt{DockingRunner}, and \texttt{DockingVirtualScreen}. A docking simulation in \texttt{pyscreener} is fully described by a \texttt{CalculationData} and an associated \texttt{CalculationMetadata}. High-level information about the simulation that is common to all docking software (e.g., target protein, docking box, name of the ligand, the paths under which inputs and outputs will be stored) is stored in the \texttt{CalculationData} object. A \texttt{CalculationMetadata} contains the set of software-specific arguments for the simulation, such as \texttt{exhaustiveness} for the AutoDock Vina family of software or parameters for SPH file preparation for DOCK6. The \texttt{pyscreener} object model separates the data from behavior by placing the responsibility of actually preparing, running, and parsing simulations inside the \texttt{DockingRunner} class. This stateless class defines methods to prepare simulation inputs, perform the simulation of the corresponding inputs, and parse the resulting output for a given \texttt{CalculationData} and \texttt{CalculationMetadata} pair. By placing this logic inside static methods rather than attaching them to the \texttt{CalculationData} object, \texttt{pyscreener} limits network data transfer overhead during task distribution. The separation also allows for the straightforward addition of new backend docking engines to \texttt{pyscreener}, as this entails only the specification of the corresponding \texttt{CalculationMetadata} and \texttt{DockingRunner} subclasses.

\texttt{pyscreener} also contains the \texttt{DockingVirtualScreen} class, which contains a template \texttt{CalculationData} and \texttt{CalculationMetadata} with which to dock each input molecule, i.e., a virtual screening protocol. The class defines a \texttt{\_\_call\_\_()} method which takes as input in SMILES strings or chemical files in any format supported by OpenBabel \cite{oboyle_open_2011} and distributes the corresponding docking simulations across the resources in the given allocation, returning a docking score for each input molecule.

\begin{figure}[b!]
    \centering
    \includegraphics[width=0.8\textwidth]{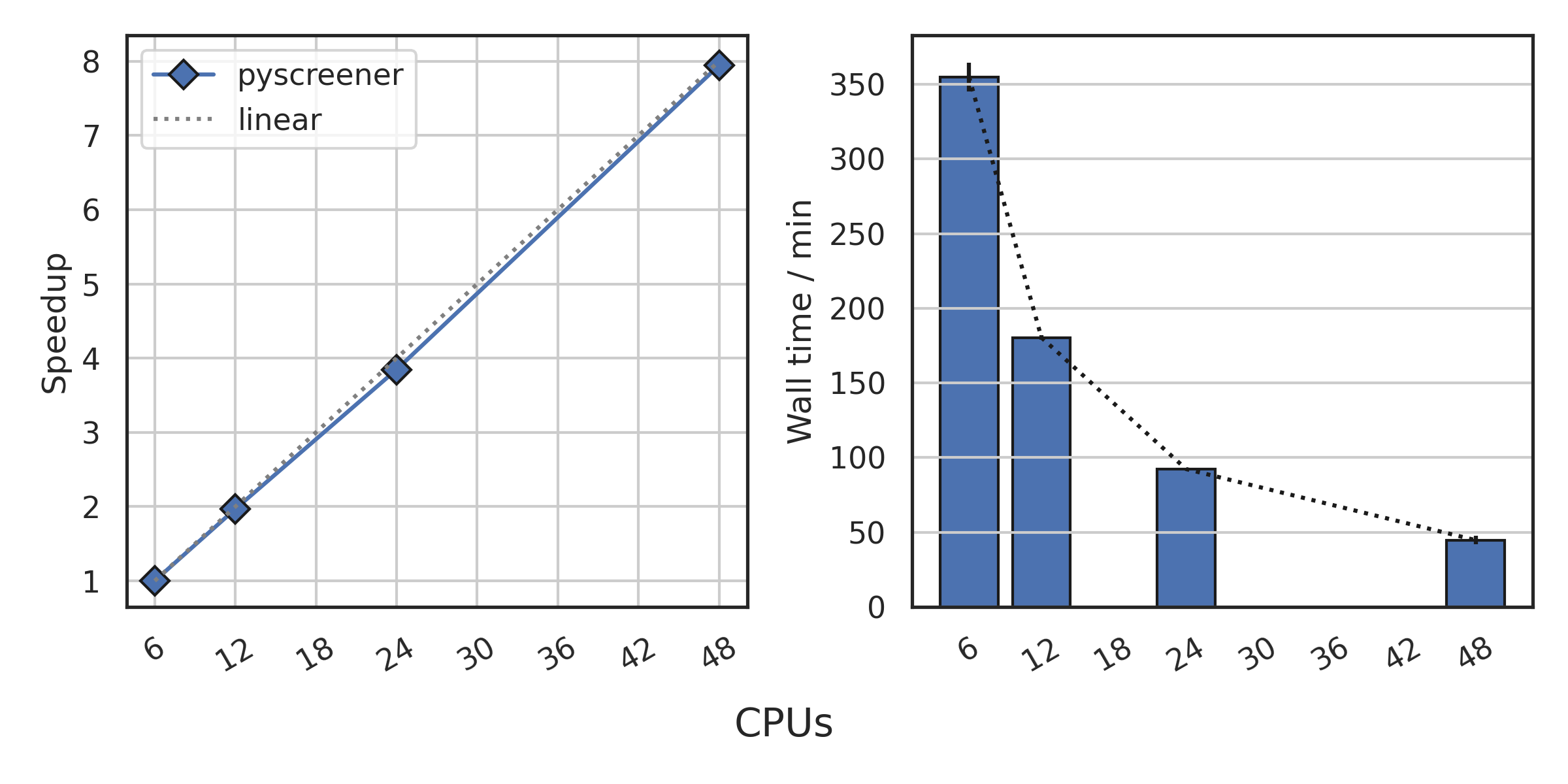}
    \caption{Wall-time of the computational docking of all 1,615 FDA-approved drugs against 5WIU using QVina over six CPU cores for a single-node setup with the specified number of CPU cores. (Left) calculated speedup. (Right) wall time in minutes. Bars reflect mean $\pm$ standard deviation over three runs.}
    \label{fig:local}
\end{figure}

\begin{figure}[t!]
    \centering
    \includegraphics[width=0.8\textwidth]{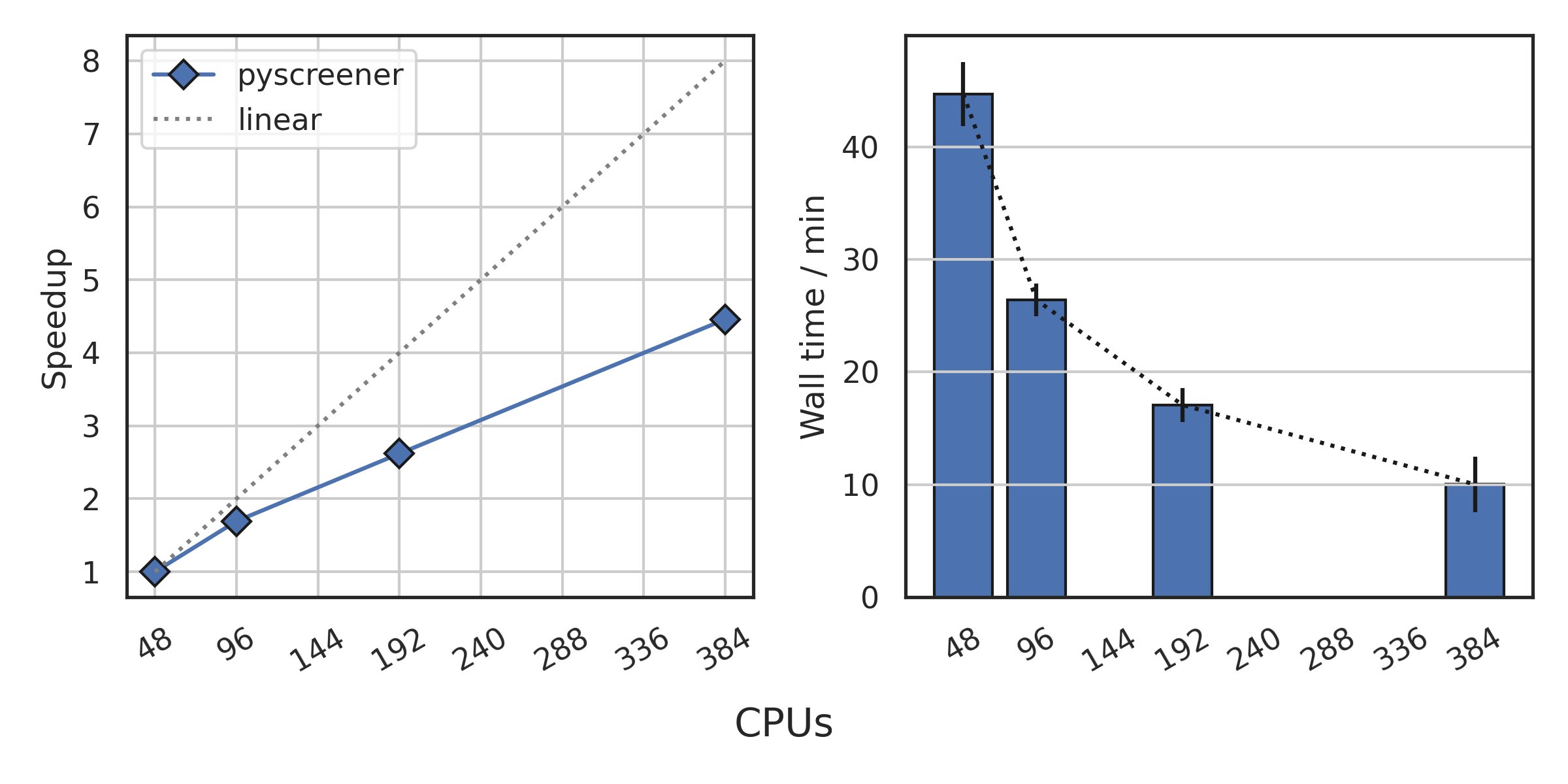}
    \caption{Wall-time of the computational docking of all 1,615 FDA-approved drugs against 5WIU using QVina over six CPU cores for setups using multiple 48-core nodes with the total number of specified CPU cores. (Left) calculated speedup. (Right) wall time in minutes. Bars reflect mean $\pm$ standard deviation over three runs.}
    \label{fig:dist}
\end{figure}

To handle task distribution, \texttt{pyscreener} relies on the \texttt{ray} library \cite{moritz_ray_2018} for distributed computation. For multithreaded docking software, \texttt{pyscreener} allows a user to specify how many CPU cores to run each individual docking simulation over, running as many docking simulations in parallel as possible for a given number of total CPU cores in the \texttt{ray} cluster. To examine the scaling behavior of \texttt{pyscreener}, we docked all 1,615 FDA-approved drugs into the active site of the D4 dopamine receptor (PDB ID 5WIU \cite{wang_d4_2017}) with QVina running over 6 CPU cores. We tested both single node hardware setups, scaling the total number of CPU cores on one machine, and multi-node setups, scaling the total number of machines. In the single-node case, \texttt{pyscreener} exhibited essentially perfect scaling \ref{fig:local} as we scaled the size of the \texttt{ray} cluster from 6 to 48 CPU cores running QVina over 6 CPU cores.

In contrast, the multi-node setup exhibits less ideal scaling \ref{fig:dist} with a measured speedup approximately 55\% that of perfect scaling. We attribute this scaling behavior to hardware-dependent network communication overhead.
Distributing a \texttt{sleep(5)} function allocated 6 CPU cores per task (to mimic a fairly quick docking simulation) in parallel over differing hardware setups
led to an approximate $2.5\%$ increase in wall-time relative to the single-node setup each time the number of nodes in the setup was doubled while keeping the total number of CPU cores the same.
Such a trend is consistent with network communication being detrimental to scaling behavior. This test also communicated the absolute minimum amount of data over the network, as there were no function arguments or return values.
When communicating \texttt{CalculationData} objects (approximately 600 bytes in serialized form) over the network, as in \texttt{pyscreener}, the drop increased to $6\%$ for each doubling of the total number of nodes.
Minimizing the total size of \texttt{CalculationData} objects was therefore an explicit implementation goal. Future development will seek to further reduce network communication overhead costs to bring \texttt{pyscreener} scaling closer to ideal scaling.

\section*{Examples}

To illustrate \texttt{pyscreener}, we consider docking benezene (SMILES string ``c1ccccc1'') against 5WIU with a docking box centered at $(-18.2, 14.4, -16.1)$ with x-, y-, and z-radii $(15.4, 13.9, 14.5)$. We may perform this docking using AutoDock Vina over 6 CPU cores via \texttt{pyscreener} like so:
\begin{lstlisting}[language=Python]
>>> import pyscreener as ps
>>> metadata = ps.build_metadata("vina")
>>> virtual_screen = ps.virtual_screen("vina", receptors=["5WIU.pdb"], center=(-18.2, 14.4, -16.1), size=(15.4, 13.9, 14.5), metadata_template=metadata, ncpu=6)
>>> scores = virtual_screen("c1ccccc1")
>>> scores
array([-4.4])
\end{lstlisting}

\noindent Alternatively, we may dock many molecules by passing a \texttt{List} of SMILES strings to the \texttt{DockingVirtualScreen}:
\begin{lstlisting}[language=Python]
>>> smis = ["c1ccccc1", "O=C(Cc1ccccc1)NC1C(=O)N2C1SC(C2C(=O)O)(C)C", "C=CCN1CCC23C4C(=O)CCC2(C1CC5=C3C(=C(C=C5)O)O4)O"]
>>> scores = virtual_screen(smis)
>>> scores.shape
(3,)
\end{lstlisting}

\noindent By default, AutoDock Vina docks molecules using an \texttt{--exhaustiveness} value of 8, but we may specify a higher number in the \texttt{metadata}:
\begin{lstlisting}[language=Python]
>>> metadata = ps.build_metadata("vina", dict(exhaustivness=32))
\end{lstlisting}

\noindent We may also utilize other docking engines in the AutoDock Vina family by specifying the \texttt{software} for Vina-type metadata. Here, we use the accelerated optimization routine of QVina for faster docking. Note that we also support \texttt{software} values of ``smina'' and ``psovina'' in addition to ``vina'' and ``qvina''.
\begin{lstlisting}[language=Python]
>>> metadata = ps.build_metadata("vina", dict(software="qvina"))
\end{lstlisting}

\noindent It is also possible to dock molecules using DOCK6 in \texttt{pyscreener}. To do this, we must first construct DOCK6 metadata and specify that we are creating a DOCK6 virtual screen (note that DOCK6 is not multithreaded and thus does not benefit from being assigned multiple CPU cores per task):
\begin{lstlisting}[language=Python]
>>> metadata = ps.build_metadata("dock")
>>> virtual_screen = ps.virtual_screen("dock", receptors=["5WIU.pdb"], center=(-18.2, 14.4, -16.1), size=(15.4, 13.9, 14.5), metadata_template=metadata)
>>> scores = virtual_screen("c1ccccc1")
>>> scores
array([-12.35])
\end{lstlisting}

\section*{Acknowledgements}
The authors thank Keir Adams and Wenhao Gao for providing feedback on the preparation of this paper and the `pyscreener` code. The timing tests in this paper were run on the FASRC Cannon cluster supported by the FAS Division of Science Research Computing Group at Harvard University. The authors also acknowledge the MIT SuperCloud and Lincoln Laboratory Supercomputing Center for providing HPC and consultation resources that have contributed to the research results reported within this paper. This work was funded by the MIT-IBM Watson AI Lab.
\printbibliography

\end{document}